\documentclass[12pt,a4paper]{article}
\usepackage{xcolor}
\usepackage{jheppub}
\usepackage{epstopdf}
\usepackage{graphicx}
\usepackage{epsfig}
\usepackage{dcolumn}  
\usepackage{bm}    
\usepackage{amssymb} 
\usepackage{amsmath,bm}
\usepackage{amsfonts}    
\usepackage{dsfont}
\usepackage{slashed}  
\usepackage{youngtab}
\usepackage{tikz}
\usepackage{pgf}
\usepackage{mathrsfs}
\usepackage{mathtools}
\usetikzlibrary{arrows}
\usepackage[mathscr]{euscript}
\usepackage{epsfig}
\usepackage{pdfpages}
\usepackage{verbatim}
\usepackage{tensor}

\newdimen\tableauside\tableauside=1.0ex
\newdimen\tableaurule\tableaurule=0.4pt
\newdimen\tableaustep
\def\phantomhrule#1{\hbox{\vbox to0pt{\hrule height\tableaurule width#1\vss}}}
\def\phantomvrule#1{\vbox{\hbox to0pt{\vrule width\tableaurule height#1\hss}}}
\def\sqr{\vbox{%
		\phantomhrule\tableaustep
		\hbox{\phantomvrule\tableaustep\kern\tableaustep\phantomvrule\tableaustep}%
		\hbox{\vbox{\phantomhrule\tableauside}\kern-\tableaurule}}}
\def\squares#1{\hbox{\count0=#1\noindent\loop\sqr
		\advance\count0 by-1 \ifnum\count0>0\repeat}}
\def\tableau#1{\vcenter{\offinterlineskip
		\tableaustep=\tableauside\advance\tableaustep by-\tableaurule
		\kern\normallineskip\hbox
		{\kern\normallineskip\vbox
			{\gettableau#1 0 }%
			\kern\normallineskip\kern\tableaurule}%
		\kern\normallineskip\kern\tableaurule}}
\def\gettableau#1 {\ifnum#1=0\let\next=\null\else
	\squares{#1}\let\next=\gettableau\fi\next}

\allowdisplaybreaks[1]

\def\be{\begin{equation}}
\def\ee{\end{equation}}
\newcommand{\abs}[1]{\left| #1 \right|}

\title{Supersymmetric $\mathrm{AdS}_3$ supergravity backgrounds and holography}

\author{Lorenz Eberhardt} 
\affiliation{Institut f\"ur Theoretische Physik, ETH Zurich, \\
\hspace*{0.3cm}CH-8093 Z\"urich, Switzerland}
\affiliation{School of Natural Sciences, Institute for Advanced Study, \\
\hspace*{0.3cm}Princeton, NJ 08540, USA}
\emailAdd{eberhardtl@itp.phys.ethz.ch}

\abstract{We analyse the conditions for $\mathrm{AdS}_3 \times \mathcal{M}_7$ backgrounds with pure NS-NS flux to be supersymmetric. We classify all $\mathcal{N}=(2,2)$ solutions where $\mathcal{M}_7$ satisfies the stronger condition of being a $\mathrm{U}(1)$-fibration over a K\"ahler manifold. We compute the BPS spectrum of all the backgrounds in this classification. We assign a natural dual CFT to the backgrounds and confirm that the BPS spectra agree, thus providing evidence in favour of the proposal.
}

\makeatletter
\g@addto@macro\bfseries{\boldmath}
\makeatother

\begin{document}

\maketitle

\section{Introduction}
$\mathrm{AdS}_3$ supergravity backgrounds provide an interesting playground to explore the $\mathrm{AdS}/\mathrm{CFT}$ correspondence \cite{Maldacena:1997re}. The case of $\mathrm{AdS}_3$ is particularly tractable, since the dual CFT's are often exactly solvable. Moreover, there is an abundance of two-dimensional CFTs --- a fact which is reflected in the richness of the corresponding supergravity backgrounds. On the other hand, this makes it hard to classify $\mathrm{AdS}_3$-backgrounds.

However, one can make some progress when imposing a sufficient amount of supersymmetry. There are only three type IIB NS-NS flux $\mathcal{N}=(4,4)$ backgrounds: $\mathrm{AdS}_3 \times \mathrm{S}^3 \times \mathcal{M}_4$, where $\mathcal{M}_4=\mathbb{T}^4$, $\mathrm{K3}$ or $\mathrm{S}^3 \times \mathrm{S}^1$. The former two possibilities support the small $\mathcal{N}=(4,4)$ algebra \cite{Aharony:1999ti}, whereas the latter supports the large $\mathcal{N}=(4,4)$ algebra \cite{Elitzur:1998mm, Gukov:2004ym, Eberhardt:2017fsi, Eberhardt:2017pty, Baggio:2017kza}. These backgrounds have very simple properties: They can be supported exclusively by NS-NS fields, exclusively by R-R fields or mixed flux. Those possibilities are related by the $\mathrm{SL}(2,\mathbb{R})$-symmetry of type IIB supergravity. NS-NS backgrounds allow for a simple string world-sheet description \cite{Maldacena:2000hw}, while R-R backgrounds are believed to have the simplest dual CFT's \cite{Maldacena:1997re}. 

Moving on to less supersymmetry, there is one known $\mathcal{N}=(4,2)$ background \cite{Donos:2014eua}. However this background is much more involved, in particular it requires all form-fields to be turned on.

For a smaller amount of supersymmetry, classifications are more difficult. When allowing five-form flux only, the geometry is very restricted. For constant axio-dilaton and $\mathcal{N}=(2,0)$, the internal manifold is a $\mathrm{U}(1)$-fibration over a K\"ahler manifold \cite{Kim:2005ez}, which satisfies some additional curvature constraints. This was generalized in \cite{Couzens:2017way} to varying axio-dilaton using an F-theory language. In particular, it was found that for $\mathcal{N}=(4,0)$-supersymmetry, the most general geometry in this case is $\mathrm{AdS}_3 \times \mathrm{S}^3/\mathbb{Z}_M \times \mathrm{CY}_3$. Here $\mathrm{CY}_3$ is an elliptically fibred Calabi-Yau three-fold, where the complex structure of the fiber is given by the axio-dilaton. 

One other direction was recently explored in \cite{Wulff:2017zbl}. There, all symmetric space solutions of type IIB supergravity were analysed. Interestingly, it was found that all $\mathrm{AdS}_3$ symmetric space solutions are related via T-duality to one of the aforementioned backgrounds $\mathrm{AdS}_3 \times \mathrm{S}^3 \times \mathcal{M}_4$ with $\mathcal{N}=(4,4)$ supersymmetry. Furthermore, all these backgrounds have $(4,4)$ or $(4,0)$ supersymmetry. Thus, both the symmetric space $\mathcal{N}=(4,0)$ solutions and the $\mathcal{N}=(4,0)$ solutions with five-form flux only are related to the known $\mathcal{N}=(4,4)$ solutions by either T-duality or are quotients thereof.\footnote{Note that this is strictly speaking only true for $\mathcal{M}_4=\mathbb{T}^4$, for K3 we cannot perform T-dualities to relate the D3-brane system of \cite{Couzens:2017way} to the D1-D5 system considered here.}

In this note we enlarge the classification result to incorporate $\mathcal{N}=(2,2)$ supersymmetry. This amount of supersymmetry is particularly attractive, since one still has good control over protected quantities like the BPS spectrum and the elliptic genus. This allows one in particular to determine the dual CFT. While previous structural results in this direction have been obtained \cite{Berenstein:1999gj, Giveon:1999jg}, they are quite indirect: A background $\mathrm{AdS}_3 \times \mathcal{M}_7$ enjoys $\mathcal{N}=(2,2)$ supersymmetry when $\mathcal{M}_7$ is a $\mathrm{U}(1)$-fibration over a six-dimensional space which is the target of an $\mathcal{N}=(2,2)$ sigma-model. This result was obtained from string theory. This result is conceptually nice, but provides little intuition on the geometry of $\mathcal{M}_7/\mathrm{U}(1)$ or on the dual CFT.

Recently, some $\mathcal{N}=(2,2)$ backgrounds were discussed \cite{Datta:2017ert}, mostly from a string perspective. They involved taking specific orbifolds of $\mathrm{AdS}_3 \times \mathrm{S}^3 \times \mathbb{T}^4$. Most of the orbifold singularities cannot be resolved, which renders the backgrounds non-geometric. However, to all of the string models, a dual CFT can be associated. Comparing BPS spectra and elliptic genera yields very non-trivial evidence for the proposal. To our knowledge, very few $\mathcal{N}=(2,2)$-backgrounds prior to \cite{Datta:2017ert} were known, which demonstrates the scarcity of these backgrounds.\footnote{In \cite{Maldacena:2000mw}, $\mathcal{N}=(2,2)$ backgrounds were constructed by compactifying $\mathcal{N}=4$ SYM on a Riemann surface with a suitable twisting. The dual supergravity background flows in the IR to a $\mathcal{N}=(2,2)$ supergravity background.}

\bigskip
In this paper, we revisit the problem from the point of view of supergravity to understand the scarcity of the $\mathcal{N}=(2,2)$ backgrounds. Motivated by the string computation of \cite{Berenstein:1999gj, Giveon:1999jg, Datta:2017ert} and to keep the calculation manageable, we consider the case of pure NS-NS flux and constant dilaton. This subsector of IIB supergravity is also known as heterotic supergravity (with trivial gauge group). Via $\mathrm{SL}(2,\mathbb{R})$-symmetry, this also finds the pure R-R solutions. The full U-duality group can typically also generate solutions with five-form flux turned on. 

We investigate $\mathcal{N}=(2,2)$ backgrounds. First, we review the constraints imposed by the existence of a Killing spinor in thos backgrounds. Supersymmetry requires the manifold $\mathcal{M}_7$ to be a local $\mathrm{U}(1)$-fibration over a \textit{conformally balanced} KT manifold \cite{Strominger:1986uh, LopesCardoso:2002vpf} $\mathcal{M}_6$. $\mathcal{M}_6$ fails to be K\"ahler, as the fundamental $(1,1)$-form $J$ of the manifold is not necessarily closed, but $J \wedge J$ is. This result was already obtained in \cite{Beck:2015gqa} using heterotic supergravity. There are known backgrounds which happen to be K\"ahler, namely the small $\mathcal{N}=(4,4)$ backgrounds we mentioned above. In contrast, the large $\mathcal{N}=(4,4)$ background corresponds to a conformally balanced manifold. 

It is interesting that the dual CFTs of the small $\mathcal{N}=(4,4)$ cases have been known for such a long time, whereas progress in the large $\mathcal{N}=(4,4)$ case was made only very recently \cite{Eberhardt:2017pty}. The failure of $\mathcal{M}_6$ to be K\"ahler is just another incarnation of the difficulty. To make progress in a classification, we will consider the case when $\mathcal{M}_6$ is a K\"ahler manifold. In that case, we succeed in completely classifying the possible internal manifolds $\mathcal{M}_7$. The main result will be that all are quotients of $\mathrm{S}^3 \times \mathbb{T}^4$ or $\mathrm{S}^3 \times \mathrm{K3}$.

We will subsequently classify all possible quotients of $\mathrm{S}^3 \times \mathbb{T}^4$ and $\mathrm{S}^3 \times \mathrm{K3}$ leading to $\mathcal{N}=(2,2)$ supersymmetry. There is a unique such quotient for $\mathrm{S}^3 \times \mathrm{K3}$, whereas there are seven for $\mathrm{S}^3 \times \mathbb{T}^4$. This result has a similar flavour as the $\mathcal{N}=(4,0)$ classification results we mentioned above. Also in this case, the backgrounds are all related to $\mathcal{N}=(4,4)$ backgrounds by quotients. Note that non-abelian T-dualities typically break supersymmetry of one chirality completely and thus do not yield $\mathcal{N}=(2,2)$ backgrounds.

We will then identify the dual CFTs of these backgrounds. The main claim is that $\mathrm{AdS}_3 \times (\mathrm{S}^3 \times \mathcal{M}_4)/\mathrm{G}$ is dual to a marginal deformation of the symmetric product orbifold of $\mathcal{M}_4/\mathrm{G}$. Here, $\mathcal{M}_4$ is $\mathbb{T}^4$ or $\mathrm{K3}$ and $\mathcal{M}_4/\mathrm{G}$ is either the Enriques surface or a hyperelliptic surface. To support this claim, we compute the chiral-chiral spectrum and the chiral-antichiral spectrum in string theory and match it to the CFT calculation. The methods and ideas follow our earlier paper \cite{Datta:2017ert}, in particular one of the solutions we present appeared already in this earlier discussion.
\bigskip

This paper is organized as follows. In Section~\ref{sec:conditions}, we review the conditions imposed on the compactification manifold imposed by the existence of Killing spinors. Starting from Section~\ref{sec:kaehler}, we discuss the backgrounds which satisfy the aforementioned K\"ahler condition and classify them completely. In Section~\ref{sec:BPS}, we will compute the BPS spectra of these backgrounds, which will be matched to the proposed CFTs in Section~\ref{sec:dual CFTs}. Some more technical calculations and conventions are summarized in various appendices.
\section{Review of the conditions imposed by supersymmetry} \label{sec:conditions}
We will assume throughout this note that the background geometry is of the form
\be 
\mathrm{AdS}_3 \times \mathcal{M}_7\ .
\ee
We will consider the NS-NS sector of type IIB supergravity on this background. This is no loss of generality, since it was shown in \cite{Beck:2015gqa} from the point of view of heterotic supergravity that the warp factor is trivial in this case. $\mathcal{M}_7$ will be assumed to be compact. We will set all fields, except for the metric and the 3-form, to zero (or constant in the case of the dilaton) and we will not consider warp factors. In particular, no R-R fields are turned on.

We will review the geometry of $\mathcal{M}_7$, as determined in \cite{Beck:2015gqa}. We rederived the results using the constraints imposed by the existence of spinor bilinears \cite{Gauntlett:2002sc, Lu:1998nu, Fujii:1985bg, Gauntlett:2002fz, Martelli:2003ki, HackettJones:2004yi, Kim:2005ez, Gauntlett:2006af, Gauntlett:2006qw, Beck:2015gqa}. 

\subsection{Killing spinor equations}
In the following, big latin indices $M,N,\dots$ denote ten-dimensional indices and small latin indices $a,b,\dots $ denote $\mathcal{M}_7$ indices. The relevant part of the type IIB action in string frame is then
\be 
\mathcal{S}^{\mathrm{NS-NS}}=\int \mathrm{d}^{10}x \sqrt{-g}\mathrm{e}^{-2\Phi}\left(R-\frac{1}{12}H_{MNP}H^{MNP}+4\nabla_M \Phi \nabla^M\Phi\right)\ ,
\ee
where $R$ is the scalar curvature of the metric $g_{MN}$ and $H$ is the field strength corresponding to the two-form $B$. As advertised above, we set the dilaton $\Phi$ to a constant. When inserting the prescribed $\mathrm{AdS}_3$-part of the fields, we end up with the Einstein equation on $\mathcal{M}_7$:
\be 
R^{ab}=\frac{1}{4} H^{acd}\tensor{H}{^b_{cd}}\ . \label{Einstein equation}
\ee
It is implied by the existence of a single Killing spinor on $\mathcal{M}_7$.

On top of the equations of motions, Killing spinor equations should be satisfied to guarantee supersymmetry of the background. The dilatino Killing spinor equation on $\mathcal{M}_7$ reads
\be 
\left(H_{abc}\gamma^{abc}+\frac{12\mathrm{i}}{\ell}\right)\eta=0\ . \label{dilatino Killing spinor M7}
\ee
Here, $\ell$ is the $\mathrm{AdS}_3$-radius. Similarly, we have a gravitino Killing spinor equation:
\be 
\nabla_a \eta_{\pm}=\mp \frac{1}{8} H_{abc} \gamma^{bc}\eta_{\pm}\ . \label{M7 gravitino Killing spinor equation}
\ee
\subsection{G-structure} 
Out of a Killing spinor $\eta$, we can form the following real forms:
\begin{align} 
C&\equiv\eta^\dag \eta\ , \\
K_a&\equiv\eta^\dag \gamma_a \eta\ , \\
J_{ab}&\equiv-\mathrm{i}\, \eta^\dag \gamma_{ab} \eta\ , \\
Z_{abc}&\equiv\mathrm{i}\, \eta^\dag \gamma_{abc} \eta\ .
\end{align}
$C$ is constant and we normalize $\eta$ such that $C \equiv 1$. We may also define some complex forms:
\begin{align} 
\Phi&\equiv\eta^{\mathrm{T}} \eta\ , \\
\Omega_{abc}&\equiv\mathrm{i}\, \eta^{\mathrm{T}} \gamma_{abc}\eta\ .
\end{align}
Other forms vanish since we chose the gamma-matrices to be in the Majorana-representation.Similarly to $C$, $\Phi$ is constant. We can redefine the Killing spinor such that $\Phi \equiv 0$.

These forms satisfy Fierz identities. In particular, we have $Z=K \wedge Y$, so $Z$ is not an independent form. It follows now from the Killing spinor equations that $K_a$ is a Killing vector and exhibits $\mathcal{M}_7$ locally as a $\mathrm{U}(1)$-fibration over a manifold $\mathcal{M}_6$. We can choose coordinates such that $K^a=\partial_\psi$. We also define $B\equiv K-\mathrm{d\psi}$. After redefining $\omega\equiv\mathrm{e}^{2\mathrm{i}\, \ell^{-1}\psi}\Omega$, it turns out that the forms $J_{ab}$ and $\omega_{abc}$ live entirely on $\mathcal{M}_6$. $J$ defines a complex structure, whereas $\omega$ defines a compatible $\mathrm{SU}(3)$-structure. We have
\begin{align}
\mathrm{d}J&=\iota_K \star H\ , \\
\mathrm{d}\omega&=2\mathrm{i}\, \ell^{-1} B \wedge \omega\ .
\end{align}
We have also in particular
\be 
\mathrm{d}(J \wedge J)=0\ .
\ee
Hence $\mathcal{M}_7$ is locally a circle fibration over a conformally balanced KT manifold \cite{Beck:2015gqa}.
\section{The K\"ahler case} \label{sec:kaehler}
We will now consider the case where $\mathcal{M}_6$ is K\"ahler, which amounts to the condition $K \wedge H=0$, since then $J$ is closed. We will refer to these backgrounds in the following as `K\"ahler backgrounds'. We will be able to give a complete classification of these backgrounds.
\subsection{Further conditions imposed by $H$}
The gravitino Killing equation for $K$ reads $\mathrm{d}K=\iota_K H$, which together with $K \wedge H=0$ implies
\be 
H=K \wedge \iota_K H=K \wedge \mathrm{d}K= K \wedge \mathrm{d}B\ .
\ee
Furthermore, the form $\mathrm{d}B$ may be identified with the Ricci-form $\rho$ of the K\"ahler manifold \cite{Chiossi:2002aa, Kim:2005ez}
. Thus, $B$ is fixed by $\rho$, so in particular the $\mathrm{U}(1)$-fibration is uniquely fixed up to the addition of a parallel vectorfield.\footnote{The argument goes as follows. Since $\mathrm{d}B=\rho$, there is locally a remaining gauge freedom of $B \to B+\mathrm{d}\Lambda$ for some function $\Lambda$. Switching our view to $\mathcal{M}_7$, $K$ is fixed up to $K \to K +\mathrm{d}\Lambda$. However, $K$ is required to be Killing, so the function $\Lambda$ must also satisfy
\be 
\nabla_{(a}\nabla_{b)}\Lambda=\nabla_a\nabla_b\Lambda=0\ .
\ee
Thus, $\mathrm{d}\Lambda$ is a parallel vectorfield. This is quite restrictive, and we easily see that there are no more conditions.
} Thus, we have
\be 
H=K \wedge \rho\ . \label{H expressed as K wedge rho}
\ee
The Ricci-form is also of type $(1,1)$.
Finally, we also need to require $H$ to be closed, since this does not follow from the conditions we have imposed so far. Since $\mathrm{d}K=\rho$, we need to require
\be 
\rho \wedge \rho =0\ .
\ee
Finally, we remark that the norm of the Ricci form of $\mathcal{M}_6$ is constant. Indeed, the norm of $\rho$ equals the norm of $H$, which in turn is constant:
\be 
H^2=\rho^2=4\ell^{-2}\ . \label{rho constant norm}
\ee
\subsection{Reducing to a local product of K\"ahler-Einstein spaces}
Let us diagonalize $\rho$ when viewed as an endomorphism acting on the tangent space. Since $\rho$ is a $(1,1)$-form, it has the standard form
\be 
\rho=\sum_{i=1}^3\lambda_i \mathrm{d}z_i \wedge \mathrm{d} \overline{z}_i\ ,
\ee
for some choice of coordinates $z_1$, $z_2$ and $z_3$. The condition $\rho \wedge \rho=0$ implies then that only one of the three eigenvalues $\lambda_1$, $\lambda_2$ and $\lambda_3$ can be non-zero, say $\lambda\equiv \lambda_1$. Moreover, since the norm of $\rho$ has to be constant by \eqref{rho constant norm}, we conclude that $\lambda$ is constant. Thus, the Ricci-tensor has only constant non-negative eigenvalues.

Now we can use the theorem proved in \cite{Apostolov:2001aa} to conclude that $\mathcal{M}_6$ is locally the product of two K\"ahler-Einstein manifolds. One manifold is of dimension 2 with positive curvature, the other of dimension 4 with vanishing curvature, i.e.~a Calabi-Yau manifold.
In other words, we have demonstrated that $\mathcal{M}_6$ is of the form
\be 
\mathcal{M}_6 \cong (\mathrm{S}^2 \times \mathrm{CY}_2)/\mathrm{G}\ ,
\ee
where $\mathrm{G}$ is some group of isometries preserving all the relevant structures. Furthermore, for $\mathcal{M}_6$ to be smooth, $\mathrm{G}$ has to act freely. Here we used that $\mathrm{S}^2$ is the only two-dimensional K\"ahler-Einstein manifold with positive K\"ahler-Einstein constant. Finally, we can change our viewpoint again to $\mathcal{M}_7$. Namely, since the $\mathrm{U}(1)$-fibration is parametrized by $\rho$, it is actually a fibration over the two-sphere $\mathrm{S}^2$, up to the aforementioned ambiguity of adding a parallel vectorfield. To eliminate this possibility, we make a case-by-case analysis. There are two possible choices for $\mathrm{CY}_2$:
\begin{enumerate}
\item $\mathrm{CY}_2=\mathbb{T}^4$. Choosing the coordinates in the appropriate fashion, we can assume that we have a $\mathrm{U}(1)$-fibration over $\mathrm{S}^2 \times \mathrm{S}^1$. However, through the canonical isomorphism $\mathrm{H}^2(\mathrm{S}^2 \times \mathrm{S}^1;\mathbb{Z}) \cong \mathrm{H}^2(\mathrm{S}^2;\mathbb{Z})$ and using the fact that the second cohomology group classifies $\mathrm{U}(1)$-bundles, we see that all $\mathrm{U}(1)$-fibrations are actually only over $\mathrm{S}^2$. Thus, in this case the freedom of adding a parallel vectorfield is trivial.
\item $\mathrm{CY}_2=\mathrm{K3}$. Since $\mathrm{K3}$ has no parallel vectorfields, this question does not arise.
\end{enumerate}
As we have demonstrated above, the fibration cannot be trivial, thus it must be the Hopf-fibration over $\mathrm{S}^2$. This can also be seen explicitly, since we have now uniquely fixed $K$.

Thus, we finally conclude that $\mathcal{M}_7$ is a finite quotient of $\mathrm{S}^3 \times \mathbb{T}^4$ or $\mathrm{S}^3 \times \mathrm{K3}$.  It is very well-known that $\mathcal{M}_7 \cong \mathrm{S}^3 \times \mathrm{CY}_2$ leads to $\mathcal{N}=(4,4)$ supersymmetry, so the group action has to be non-trivial. This also finally demonstrates that the requirements we imposed were sufficient, since $\mathrm{S}^3 \times \mathrm{CY}_2$ satisfies the supergravity equations.
\subsection{The case of $\mathcal{M}_7 \cong \mathrm{S}^3 \times \mathrm{CY}_2$}
To continue, it is advantageous to have a good understanding of $\mathcal{M}_7 \cong \mathrm{S}^3 \times \mathbb{T}^4$, so we review here the background following \cite{Datta:2017ert}. We have $H \propto \mathrm{vol}_{\mathrm{S}^3}$ in this case. Thus the above gravitino Killing spinor reduces to the standard one on $\mathrm{S}^3$, while on $\mathbb{T}^4$ we are searching for parallel Killing spinors. $H_{abc}\gamma^{abc}$ commutes with all gamma-matrices on $\mathrm{S}^3$, but anticommutes with all on $\mathbb{T}^4$. Thus, the dilatino spinor equation imposes a definite chirality on $\mathbb{T}^4$.

It is a mathematical fact that Killing spinors with non-vanishing Killing constant are in one-to-one correspondence with parallel Killing spinors on the Riemannian cone \cite{Baer:1992aa}. The chirality of the spinor on the cone translates into the sign of the Killing constant. For the case of $\mathrm{S}^3$, its Riemannian cone is $\mathbb{R}^4$, so the problem simply reduces to finding parallel Killing spinors on $\mathbb{R}^4 \times \mathbb{T}^4$. In addition, they have to satisfy the dilatino Killing spinor equation.

Now we can count the number of Killing spinors of type IIB on this background. We are considering first one ten-dimensional Majorana-Weyl Killing spinor. Standard counting tells us that $\mathbb{R}^4 \times \mathbb{T}^4$ possesses $2 \times 2^4=32$ parallel Dirac spinors. Half of them have the correct chirality and hence the correct sign of the Killing constant on $\mathrm{S}^3 \times \mathbb{T}^4$.\footnote{This sign depends on the ten-dimensional Killing spinor we are considering.} Furthermore, only half of those satisfy the chirality constraint on $\mathbb{T}^4$ and hence obey the dilatino Killing-spinor equation. Similarly, there are $\tfrac{1}{2} \times 2 \times 2^2=4$ Dirac Killing spinors on $\mathrm{AdS}_3$ with the correct sign of the Killing constant. Putting these Killing spinors together gives $4 \times 8=32$ ten-dimensional Dirac Killing spinors. See e.g.~\cite{Lu:1998nu} on how to combine the two spinors into a ten-dimensional spinor. Now we have to impose also that the ten-dimensional Killing spinor is Majorana-Weyl, which gives then only $\tfrac{1}{2} \times \tfrac{1}{2}\times 32=8$ Majorana-Weyl Killing spinors. The same holds true for the other ten-dimensional Killing spinor. Thus, we have in total 8 left-moving supersymmetries and 8 right-moving supersymmetries. Hence this leads to $\mathcal{N}=(4,4)$ supersymmetry. 

The case of $\mathcal{M}_7 \cong \mathrm{S}^3 \times \mathrm{K3}$ works in essentially the same way. Here the Killing spinors already have a definite chirality on $\mathrm{K3}$, since the holonomy group is $\mathrm{SU}(2)$. Thus, the dilatino equation is superfluous, and the same argument as before yields again $\mathcal{N}=(4,4)$ supersymmetry.
\subsection{All K\"ahler possibilities}
We will now systematically explore all possibilities of quotients of $\mathrm{S}^3 \times \mathrm{CY}_2$ which preserve $\mathcal{N}=(2,2)$ supersymmetry. For this, let $\mathrm{G} \subset \mathrm{Isom}(\mathrm{S}^3 \times \mathrm{CY}_2)$ some group of isometries by which we want to quotient. Obviously,
\be 
\mathrm{Isom}(\mathrm{S}^3 \times \mathrm{CY}_2) \cong \mathrm{O}(4) \times \mathrm{Isom}(\mathrm{CY}_2)\ ,
\ee
so we may look on the action on $\mathrm{S}^3$ and $\mathrm{CY}_2$ separately. To keep things simple, we only consider actions which are orientable on both $\mathrm{S}^3$ and on $\mathrm{CY}_2$, since otherwise supersymmetry turns out to be completely broken. The spin double cover of $\mathrm{SO}(4)$ is $\mathrm{SU}(2)\times \mathrm{SU}(2)$, where the two factors correspond to the two different chiralities. Clearly the group action has to be non-trivial in both factors, since otherwise we would not obtain $\mathcal{N}=(2,2)$ supersymmetry. 

Let us first consider a cyclic subgroup of the group of isometries. We can choose the coordinates in such a way that the $\mathrm{S}^3$-part (or rather its lift to the spin-bundle) lies in the standard Cartan-subalgebra $\mathrm{U}(1) \times \mathrm{U}(1) \subset \mathrm{SU}(2) \times \mathrm{SU}(2)$. Now we claim that the $\mathrm{S}^3$-action actually lies in the diagonal or anti-diagonal combination of $\mathrm{U}(1) \times \mathrm{U}(1)$, or it lies entirely in one of the $\mathrm{U}(1)$'s. If this would not be the case, the group element had four different eigenvalues on the spinor representation $(\mathbf{2},\mathbf{1}) \oplus (\mathbf{1},\mathbf{2})$. There is an additional phase which can be produced by the action of the group element on $\mathrm{CY}_2$, but it is the same for all four states in the representation. Thus, at least three states get projected out, and we remain at most with $(2,0)$-supersymmetry. The case where the isometries lie completely in one of the $\mathrm{U}(1)$'s can be discarded, since it destroys one chirality of spinors completely and leaves the other untouched. Thus, it is associated with $(4,0)$-supersymmetry, as discussed in \cite{Kutasov:1998zh}. Without loss of generality, we may assume that the cyclic subgroup lies in the diagonal $\mathrm{U}(1) \subset \mathrm{U}(1) \times \mathrm{U}(1)$. This however means in the $\mathrm{SO}(4)$-language that the action on $\mathrm{S}^3$ is given by a rotation. In particular it has a fixed point.

Each group element has to act fix-point free on $\mathrm{S}^3$ or on $\mathrm{CY}_2$ for the quotient space to be smooth. We have however seen above that each group element has to act with fixed points on $\mathrm{S}^3$. Consequently the action on $\mathrm{CY}_2$ must be free. This in turn implies that the quotient space $\mathrm{CY}_2/\mathrm{G}$ is a Calabi-Yau manifold in the weak sense. This means that it is only required to have a vanishing first Chern-class in real cohomology, but not in integer cohomology. As a consequence, these manifolds are actually not spin manifolds --- only the complete $\mathcal{M}_7$ will be a spin manifold. This is extremely restrictive and these quotients are all classified by mathematicians. A standard reference is \cite{Barth:2004aa}. There are two classes, belonging to $\mathbb{T}^4$ and K3, respectively. 

K3 has a unique such quotient, called the Enriques surface. We will denote the Enriques surface by $\mathrm{ES}$. The quotient group is a $\mathbb{Z}_2$. The Enriques surface has Hodge-diamond
\be 
\begin{tabular}{ccccc}
& & 1 & & \\
& 0 & & 0 & \\
0 & & 10 & & 0 \\
& 0 & & 0 & \\
& & 1 & &
\end{tabular}\ . \label{Enriques Hodge diamond}
\ee

$\mathbb{T}^4$ has a family with seven members of such quotients, which go by the name of (irregular) hyperelliptic surface.\footnote{Not to be confused with hyperelliptic Riemann surface.} They are also called bi-elliptic surfaces, since they admit an elliptic fibration over an elliptic curve. Thus, they are best viewed as a finite quotient of a product of an elliptic curve $E=\mathbb{C}/\Gamma$ with an elliptic curve $C \cong \mathrm{S}^1 \times \mathrm{S}^1$. We will denote them generically by HS and write e.g.~$\mathrm{HS}_{\text{b 2)}}$ to indicate a specific one. They have all the Hodge-diamond
\be 
\begin{tabular}{ccccc}
& & 1 & & \\
& 1 & & 1 & \\
0& & 2 & & 0\\
& 1 & & 1 & \\
& & 1 & &
\end{tabular}\ . \label{hyperelliptic Hodge diamond}
\ee
For the convenience of the reader, we have listed the different possibilities for the group actions in Table~\ref{tab:hyperelliptic surfaces}.  Further properties of surfaces are presented in Appendix~\ref{app:surfaces}.
\begin{table}
\begin{center}
\begin{tabular}{llll}
Type & $\Gamma$ & $G$ & Action of $G$ on $E$ \\
\hline
a 1) & arbitrary & $\mathbb{Z}_2$ & $ \lambda \mapsto -\lambda$ \\
a 2) & arbitrary & $\mathbb{Z}_2 \oplus \mathbb{Z}_2$ & $\lambda \mapsto -\lambda$ \\
& & & $ \lambda \mapsto \lambda+\mu$ with $-\mu=\mu$ \\
b 1) & $\mathbb{Z} \oplus \mathbb{Z}\omega$ & $\mathbb{Z}_3$ & $\lambda \mapsto \omega \lambda$ \\
b 2) & $\mathbb{Z} \oplus \mathbb{Z}\omega$ & $\mathbb{Z}_3\oplus \mathbb{Z}_3$ & $\lambda \mapsto \omega \lambda$ \\
& & & $\lambda \mapsto \lambda+\mu$ with $\omega\mu=\mu$ \\
c 1) & $\mathbb{Z} \oplus \mathbb{Z}\mathrm{i}$ & $\mathbb{Z}_4$ & $\lambda \mapsto \mathrm{i}\, \lambda$ \\
c 2) & $\mathbb{Z} \oplus \mathbb{Z}\mathrm{i}$ & $\mathbb{Z}_4\oplus \mathbb{Z}_2$ & $\lambda \mapsto \mathrm{i}\, \lambda$ \\
& & & $\lambda \mapsto \lambda+\mu$ with $\mathrm{i}\, \mu=\mu$\\
d) & $\mathbb{Z} \oplus \mathbb{Z}\omega$ & $\mathbb{Z}_6$ & $\lambda \mapsto -\omega \lambda$ 
\end{tabular}
\end{center}
\caption{The classification of bi-elliptic surfaces. $\omega$ is a cubic unit root. This table is taken from \cite{Barth:2004aa}.} \label{tab:hyperelliptic surfaces}
\end{table}
As a last step, we have to determine the corresponding actions of the quotient groups on $\mathrm{S}^3$. For the Enriques surface, this is immediate, since we argued before that every group element acting on $\mathrm{S}^3$ has to lie in the diagonal or anti-diagonal $\mathrm{U}(1)$. As there is only one non-trivial group element, it can either act trivially or by a rotation by $\pi$ around one axis. The former is not possible, since then $\mathcal{M}_7 \cong \mathrm{S}^3 \times \mathrm{ES}$, which is not a spin-manifold. Thus, the action on $\mathrm{S}^3$ is uniquely determined. The eigenvalues of the $\mathrm{U}(1) \times \mathrm{U}(1)$ on the spin representation $(\mathbf{2},\mathbf{1}) \oplus (\mathbf{1},\mathbf{2})$ are given by $(\mathrm{i},-\mathrm{i})$ for both the left and right chirality. Thus, precisely half of the left and right chirality Killing spinors survive the projection, since the eigenvalues on the Enriques surface are either $\mathrm{i}$ or $-\mathrm{i}$, depending on the choice of the lift of the group action to the spinor-bundle.

We can similarly argue for the hyperelliptic surfaces a 1) and b 1), where the group actions are given by a rotation by $\pi$ and $2\pi/3$, respectively. For the hyperelliptic surfaces c 1) and d 1), the same argument reveals again that the group actions are given by rotations. However the angle is no longer uniquely fixed. Looking again at the eigenvalues of the group action on the Killing spinors, we see that the angles must be $\pi/2$ and $\pi/3$, respectively, since otherwise no supersymmetry would survive. 

Finally, we come to the surfaces of type 2). We see that the second generator acts trivially on the Killing spinors, hence to preserve supersymmetry, it also has to act trivially on $\mathrm{S}^3$. Thus, the action on $\mathrm{S}^3$ for the surfaces of type 2) are precisely the same as those for their type 1) counterparts. This completes the classification of backgrounds coming from K\"ahler geometries. We have seen that the action on $\mathrm{S}^3$ is in all cases uniquely fixed.

Moreover, the example of the hyperelliptic surface which was mentioned in \cite{Datta:2017ert} fits into this classification. It is given by the hyperelliptic surface of type a 1).

\subsection{Induced action on $\mathrm{S}^3$} \label{subsec:induced action}
In this subsection, we will see that the action on $\mathrm{S}^3$ is actually very natural. For this, we remember that the background $\mathrm{AdS}_3 \times \mathrm{S}^3 \times \mathrm{CY}_2$ supports small $\mathcal{N}=(4,4)$ superconformal symmetry. Thus, in particular, it has a spacetime $\mathrm{SU}(2) \times \mathrm{SU}(2)$-symmetry, which is simply given by rotations of $\mathrm{S}^3$.

We now remember the $\mathrm{AdS}/\mathrm{CFT}$-correspondence for $\mathrm{AdS}_3 \times \mathrm{S}^3 \times \mathrm{CY}_2$. It states that supergravity on this background lies on the same moduli space as the infinite symmetric product CFT
\be 
\mathrm{Sym}^\infty(\mathrm{CY}_2)\ .
\ee
All $\mathrm{CY}_2$ manifolds are actually hyperk\"ahler manifolds and hence also support $\mathcal{N}=(4,4)$ superconformal symmetry. If we now act by some isometries on $\mathrm{CY}_2$, we consequently get an induced action on the $\mathrm{SU}(2)\cong \mathrm{S}^3$-current algebra the theory supports. By the AdS/CFT correspondence, we expect this action to be precisely the one we determined in the previous section by brute-force. Since the quotients of $\mathrm{CY}_2$ we are considering are still K\"ahler and Ricci-flat, they still support an $\mathcal{N}=(2,2)$ superconformal field theory. Thus, we conclude that the action on the $\mathrm{SU}(2)$-current has to leave invariant an $\mathrm{U}(1) \subset \mathrm{SU}(2)$. So the remaining group of automorphisms is only a $\mathrm{U}(1)$, in other words, the group acts by rotations on $\mathrm{S}^3$.
\section{The BPS spectra of the K\"ahler backgrounds} \label{sec:BPS}
We have established that all K\"ahler $\mathcal{N}=(2,2)$ backgrounds are of the form
\be 
\mathrm{AdS}_3 \times (\mathrm{S}^3 \times \mathrm{CY}_2)/\mathrm{G}\ ,
\ee
where a complete list of the possibilities was provided in the last Section. It is the next logical step to compute the type IIB supergravity and BPS spectra of these backgrounds. Note that even though the backgrounds are supported purely by NS-NS flux, we are now considering the full IIB supergravity spectrum, including R-R fields and fermions.

Since the backgrounds inherit many properties from their $\mathcal{N}=(4,4)$ cousins, we can use the techniques of \cite{deBoer:1998kjm}. For this, we use the fact that the states are still secretly sitting in $\mathcal{N}=(4,4)$ multiplets, but some states of the multiplets are projected out. We have collected some relevant background for this in Appendix~\ref{app:N4N2}. We have already applied a similar technique in \cite{Datta:2017ert}. We will denote by $(\mathbf{m},\mathbf{n})^\alpha$ a modified $\mathrm{SU}(2) \times \mathrm{SU}(2)$-multiplet, where $\alpha$ is a unit root of the order of the cyclic group action on $\mathrm{S}^3$. Furthermore, we denote by $(\mathbf{m},\mathbf{n})^\alpha_\mathrm{S}$ a short modified $\mathcal{N}=(4,4)$ multiplet. The refinement of the $\mathcal{N}=(4,4)$ multiplets with insertions of $\alpha$ helps us to keep track of the transformation properties under $\mathrm{G}$.
\subsection{The Enriques surface}
Let us first begin with the K3 case and the associated Enriques surface. In this case $\alpha$ is a second root of unity, since the group is $\mathbb{Z}_2$. In the following we let $\alpha$ be a formal variable satisfying $\alpha^2=1$. The action of $\mathbb{Z}_2$ on the Hodge-diamond of $\mathrm{K3}$ is
\be 
\begin{tabular}{ccccc}
& & 1 & & \\
& 0 & & 0 & \\
$\alpha$ & & $10(1+\alpha)$ & & $\alpha$ \\
& 0 & & 0 & \\
& & 1 & &
\end{tabular}\ . \label{Enriques surface modified Hodge diamond}
\ee
The invariant part is the constant part in $\alpha$, which gives the Hodge-diamond of the Enriques surface \eqref{Enriques Hodge diamond}. Following \cite{deBoer:1998kjm}, we first compactify to six dimensions and perform subsequently the Kaluza-Klein reduction on $\mathrm{S}^3$. During this procedure, we keep track of the eigenvalues of the projection and in the end we only keep invariant states. Furthermore, it will suffice to determine the bosonic field content, since the fermionic fields will be fixed by $\mathcal{N}=(2,2)$ supersymmetry. Compactifying type IIB supergravity to six dimensions, we obtain the bosonic field content indicated in Table~\ref{tab:Enriques 6d fields}. The fields come about as follows:
\begin{enumerate}
\item[(i)] Compactifying a ten-dimensional scalar yields again a scalar in six dimensions. Type IIB supergravity contains two scalars, hence this contributes two scalars.
\item[(ii)] Compactifying a ten-dimensional two-form gives the following field content in six dimensions. We have one six-dimensional two-form, which splits into a self-dual and an anti self-dual two-form. Furthermore, we obtain $b_1$ vectors, where $b_1$ is the first Betti number of the internal four-dimensional manifold. Finally, we obtain $b_2$ scalars. Type IIB has two ten-dimensional two-forms and K3 has the cohomology \eqref{Enriques surface modified Hodge diamond}. Thus, this contributes 2 self-dual two-forms, 2 anti self-dual two-forms and $2 \times(10+12\alpha)$ scalars.
\item[(iii)] We now consider the ten-dimensional self-dual four-form. Compactifying it yields one scalar, $b_1$ vectors and $\tfrac{1}{2} b_2$ two-forms. The two-forms can be either self-dual or anti self-dual, depending on the signature of the internal manifold. For the case of type IIB and K3, this yields one scalar and $5+6\alpha$ two-forms. These split into $1+2\alpha$ self-dual and $9+10\alpha$ anti self-dual forms. As required, the splitting is dictated by the signatures. K3 has signature $-16$, whereas ES has signature $-8$, see Appendix~\ref{app:surfaces}. When ignoring the $\alpha$-dependence ($\alpha=1$), there are hence 16 more anti self-dual forms (19) than self-dual forms (3). When performing the projection ($\alpha=0$), we would compactify on ES and hence have 8 more anti self-dual forms (9) than self-dual forms (1).
\item[(iv)] Finally, we compactify the metric. It yields one metric in six dimensions. Furthermore, we obtain a non-abelian gauge field realizing the isometry group of the compactification manifold. Finally, we obtain as many scalars as there are moduli in the compactification. For the present case, K3, as well as ES has a discrete isometry group and hence contributes no vectors in six dimensions. We left the number of scalars undetermined and denoted them by $\mathrm{dim}(\mathcal{M}_{\mathrm{ES}})$.
\end{enumerate}
It is not necessary to determine the dimension of the moduli space of string compactifications $\mathcal{M}_\mathrm{ES}$ from first principles --- it will also be fixed by $\mathcal{N}=(2,2)$ supersymmetry. Summing up yields then Table~\ref{tab:Enriques 6d fields}.
\begin{table}
\begin{center}
\begin{tabular}{ll}
type & number \\
\hline
scalar & $23+24\alpha+\mathrm{dim} (\mathcal{M}_\mathrm{ES})$ \\
vector & 0 \\
self-dual 2-form & $3+2\alpha$ \\
anti self-dual 2-form & $11+10\alpha$ \\
metric & 1 
\end{tabular} 
\end{center}
\caption{Six-dimensional fields after the compactificationof type IIB supergravity on the Enriques surface. We included also the number of odd fields under the projection, they can still contribute to the three-dimensional field content.}\label{tab:Enriques 6d fields}
\end{table} 

In the next step, we perform the KK-reduction on the sphere $\mathrm{S}^3$. The quotient has the effect of replacing the standard multiplets $(\mathbf{m},\mathbf{n})$ by the twisted multiplets $(\mathbf{m},\mathbf{n})^\alpha$, for more details on those consult Appendix~\ref{app:N4N2}. In this case, $\alpha$ has order two and hence we have to decide whether we replace the multiplet $(\mathbf{m},\mathbf{n})$ by $(\mathbf{m},\mathbf{n})^\alpha$ or by $\alpha (\mathbf{m},\mathbf{n})^\alpha$. The answer is simple: Even spin particles are clearly invariant under the group action on $\mathrm{S}^3$, whereas odd spin particles are not.\footnote{This can also be seen in a less hand-wavy manner. The representations we wrote down are $\mathrm{SO}(4)$-representations. A rotation by $180$ degrees can be represented by the element $\mathrm{diag}(-1,-1,1,1)$ in $\mathrm{SO}(4)$. This is in the Cartan-torus and the sign picked up under this rotation is then $(-1)^{\frac{1}{2}(\mathbf{n}-\mathbf{m})}$. Hence we conclude again that vectors receive an additional $\alpha$, whereas the other fields are invariant.\label{footnote:fixing the sign}} Thus vectors will be multiplied by an additional $\alpha$ in the end. Hence, following \cite{deBoer:1998kjm}, the three-dimensional bosonic field content is
\be 
\bigoplus_\mathbf{m} (\mathbf{m},\mathbf{m} \pm 4)^\alpha \oplus (12+16\alpha) (\mathbf{m},\mathbf{m}\pm 2)^\alpha \oplus (40+36\alpha+\mathrm{dim} (\mathcal{M}_\mathrm{ES})) (\mathbf{m},\mathbf{m})^\alpha\ .
\ee
This can be uniquely fitted into modified $\mathcal{N}=(4,4)$ multiplets as described in Appendix~\ref{app:N4N2} with the result
\be 
\bigoplus_\mathbf{m} \alpha (\mathbf{m},\mathbf{m}\pm 2)^\alpha_\mathrm{S} \oplus (12+10\alpha)(\mathbf{m},\mathbf{m})^\alpha_\mathrm{S} \ . \label{Enriques surface supergravity spectrum}
\ee
It is clear that there will be some exceptional cases for small values of $\mathbf{m}$, which we have not treated here. This fixes also uniquely the dimension of the moduli space of the compactification:
\be 
\mathrm{dim}(\mathcal{M}_\mathrm{ES})=30+28\alpha\ .
\ee
In particular, the chiral-chiral
BPS spectrum reads
\be 
\bigoplus_{\mathbf{m}=0}^\infty 12(\mathbf{m},\mathbf{m})\ . \label{Enriques surface chiral-chiral primary spectrum}
\ee
We can also extract the chiral-antichiral ring:
\be 
\bigoplus_{\mathbf{m}=0 \text{ even}}^\infty (\mathbf{m},-\mathbf{m}\pm 2) \oplus 10(\mathbf{m},-\mathbf{m}) \oplus \bigoplus_{\mathbf{m}=0 \text{ odd}}^\infty 12(\mathbf{m},-\mathbf{m})\ .\label{Enriques surface chiral-antichiral primary spectrum}
\ee
There is clearly a quite non-trivial structure in these invariants.
\subsection{The hyperelliptic surface}
We now repeat the analysis for the hyperelliptic surface. The $\mathbb{Z}_n$-action\footnote{We have not included the second $\mathbb{Z}_m$ which appears in the type 2) surface, since it acts trivial.} on the Hodge-diamond of $\mathbb{T}^4$ is now
\be 
\begin{tabular}{ccccc}
& & 1 & & \\
& $1+\alpha$ & & $1+\alpha^{-1}$ & \\
$\alpha$ & & $2+\alpha+\alpha^{-1}$ & & $\alpha^{-1}$ \\
& $1+\alpha$ & & $1+\alpha^{-1}$ & \\
& & 1 & &
\end{tabular}\ .\label{hyperelliptic surface modified Hodge diamond}
\ee
The six-dimensional field content is then determined as before and is collected in Table~\ref{tab:hyperelliptic surface 6d fields}.
\begin{table}
\begin{center}
\begin{tabular}{ll}
type & number \\
\hline
scalar & $4\alpha^{-1}+7+4\alpha+\mathrm{dim} (\mathcal{M}_\mathrm{HS})$ \\
vector & $4(\alpha^{-1}+2+\alpha)$ \\
self-dual 2-form & $3+2\alpha$ \\
anti self-dual 2-form & $2\alpha^{-1}+3$ \\
metric & 1 
\end{tabular} 
\end{center}
\caption{Six-dimensional fields after the compactification of type IIB supergravity on the hyperelliptic surface.
}\label{tab:hyperelliptic surface 6d fields}
\end{table} 
Now we can perform the KK-reduction as before. Fixing the sign is a bit trickier as before, since $\alpha$ does not necessarily square to one. However, the argument of footnote \ref{footnote:fixing the sign} still works and the prefactor of the representation $(\mathbf{m},\mathbf{n})$ is $\alpha^{\frac{1}{2}(\mathbf{n}-\mathbf{m})}$.
\begin{align}
\bigoplus_\mathbf{m} \, &\alpha^{\pm 2}(\mathbf{m},\mathbf{m}\pm 4)^\alpha \oplus \alpha^{\pm}(6\alpha^{-1}+16+6\alpha)(\mathbf{m},\mathbf{m}\pm 2)^\alpha \nonumber\\
&\qquad\oplus (14\alpha^{-1}+32+14\alpha+\mathrm{dim}(\mathcal{M}_\mathrm{HS}))(\mathbf{m},\mathbf{m})^\alpha\ .
\end{align}
Again, we can fit this uniquely into modified multiplets with the result
\be 
\bigoplus_\mathbf{m} \alpha^{\pm 1} (\mathbf{m},\mathbf{m}\pm 2)_\mathrm{S}^\alpha \oplus 2 (1+\alpha^{\pm 1})(\mathbf{m},\mathbf{m}\pm 1)_\mathrm{S}^\alpha \oplus (\alpha^{-1}+4+\alpha)(\mathbf{m},\mathbf{m})_\mathrm{S}^\alpha\ . \label{hyperelliptic surface supergravity spectrum}
\ee
Again, there are some exceptional cases at low spin. Furthermore, this tells us
\be 
\mathrm{dim}(\mathcal{M}_\mathrm{HS})=\alpha^{-2}+2\alpha^{-1}+4+2\alpha+\alpha^2\ .
\ee
From the supergravity spectrum we can now straightforwardly extract the chiral-chiral primary spectrum:
\be 
\bigoplus_\mathbf{m} 2 (\mathbf{m},\mathbf{m \pm 1}) \oplus 4 (\mathbf{m},\mathbf{m}) \ . \label{hyperelliptic surface chiral-chiral primary spectrum}
\ee
The chiral-antichiral primary spectrum is very interesting in this case. It can in particular distinguish different hyperelliptic surfaces. It is in general given by the constant part of
\begin{align} 
&\bigoplus_\mathbf{m} \alpha^{-\mathbf{m}} (\mathbf{m},-\mathbf{m}- 2)\oplus \alpha^{2-\mathbf{m}}(\mathbf{m},-\mathbf{m}+2) \oplus 2\alpha^{-\mathbf{m}}(1+\alpha) (\mathbf{m},-\mathbf{m}- 1) \nonumber\\
&\qquad\oplus 2(1+\alpha)\alpha^{1-\mathbf{m}}(\mathbf{m},-\mathbf{m}+1)\oplus \alpha^{-\mathbf{m}}(1+4\alpha+\alpha^2)(\mathbf{m},-\mathbf{m})\ . \label{hyperelliptic surface chiral-antichiral primary spectrum}
\end{align}
It has hence a periodicity in $\mathbf{m}$ of period equal to the order of the quotient group.
\section{Dual CFTs for the K\"ahler backgrounds} \label{sec:dual CFTs}
There are almost canonical candidates for dual CFTs to the K\"ahler backgrounds. First note that the Enriques surface and the hyperelliptic surfaces are the only geometric backgrounds besides $\mathbb{T}^4$ and $\mathrm{K3}$ which support an $\mathcal{N}=(2,2)$ superconformal algebra at $c=6$. It is thus very natural that the dual CFTs should in analogy to the case of $\mathbb{T}^4$ and $\mathrm{K3}$ correspond to the symmetric orbifold of the respective seed theories. This should also work, since we have argued in Section~\ref{subsec:induced action} that we have identified the same group actions on both sides of the small $\mathcal{N}=(4,4)$ dualities.

\medskip
We hence propose that type IIB supergravity on the supergravity backgrounds we analysed above lies on the same moduli space as the symmetric orbifolds
\be 
\mathrm{Sym}^\infty (\mathrm{ES})\ , \quad \mathrm{Sym}^\infty(\mathrm{HS})
\ee
of the Enriques surface and the corresponding hyperelliptic surface, respectively.\footnote{We expect that this correspondence continues to hold for a finite number of copies, where the CFT should be dual to a string theory on the respective background. This is in the spirit of what was found in \cite{Datta:2017ert} from the point of view of string theory.} The same proposal was made in \cite{Datta:2017ert} for the first of the hyperelliptic surfaces, so this is the natural generalization of the idea presented there.

To support the claim, we will show in this section that the chiral-chiral and chiral-antichiral primary spectrum we calculate from these CFTs agree with the ones we computed in the previous section.
\subsection{The DMVV-formula}
Denote by $Z(z|\tau)$ the partition function of the seed theory $\mathrm{ES}$ or $\mathrm{HS}$ with the insertion of $(-1)^\mathrm{F}$:
\be 
Z(z|\tau)=\mathrm{tr} \left((-1)^{\mathrm{F}}y^{J_0}\bar{y}^{\bar{J}_0}q^{L_0}q^{\bar{L}_0}\right)\ .
\ee
Here, we included a chemical potential for the $\mathrm{U}(1)$-charges. As usual,
\be 
q=\mathrm{e}^{2\pi\mathrm{i}\, \tau}\ , \quad y=\mathrm{e}^{2\pi\mathrm{i}\,  z}\ .
\ee
No holomorphicity on $\tau$ or $z$ is assumed. We add a subscript `NSNS' or `RR' to indicate whether the trace is taken in the NS-NS sector or in the R-R sector. We add a superscript $N$ to refer to the symmetric orbifold theory with $N$ copies. As one can see from the definition of the partition function, we suppressed ground state energies. We write
\be 
Z_\mathrm{RR}(z|\tau)=\sum_{m,\ell} c(m,\bar{m},\ell,\bar{\ell})q^m \bar{q}^{\bar{m}} y^\ell \bar{y}^{\bar{\ell}}\ . \label{definition cs}
\ee
In \cite{Dijkgraaf:1996xw} and \cite{Maldacena:1999bp}, a formula was given for the partition function of the symmetric orbifold:
\be 
\sum_{N=0}^\infty p^N Z_\mathrm{RR}^N(z|\tau)=\prod_{n=1}^\infty\prod_{\ell,\, m,\, \bar{\ell},\, \bar{m}} \frac{1}{(1-p^nq^m \bar{q}^{\bar{m}}y^\ell \bar{y}^{\bar{\ell}})^{c(nm,n\bar{m},\ell,\bar{\ell})}}\ . 
\ee
It is convenient to let this formula flow to the NS-NS sector:
\be 
\sum_{N=0}^\infty p^N Z_\mathrm{NSNS}^N(z|\tau)=\prod_{n=1}^\infty\prod_{\ell,\, m,\, \bar{\ell},\, \bar{m}} \frac{1}{(1-p^nq^m \bar{q}^{\bar{m}}y^\ell \bar{y}^{\bar{\ell}})^{c\left(n\left(m-\ell\right),n\left(\bar{m}-\bar{\ell}\right),\ell-\tfrac{n}{2},\bar{\ell}-\tfrac{n}{2}\right)}}\ .\label{DMVV NSNS sector}
\ee
We note that the right hand side of \eqref{DMVV NSNS sector} contains exactly one factor of the form $(1-p)^{-1}$. Following the argument of \cite{deBoer:1998us}, we can extract $Z^\infty_\mathrm{NSNS}(z|\tau)$ as follows. The right hand side of \eqref{DMVV NSNS sector} is of the form
\be 
\frac{1}{1-p}\sum_{i=0}^\infty x_i p^i=\sum_{i=0}^\infty\sum_{j=0}^i x_j p^i\ .
\ee
Hence, we can extract $Z^\infty_\mathrm{NSNS}(z|\tau)=\sum_{j=0}^\infty x_j$ by omitting the factor of $(1-p)^{-1}$ and setting $p=1$. We will indicate the fact that this factor is omitted by a prime in the product. Thus, we have
\be 
Z_\mathrm{NSNS}^\infty(z|\tau)=\prod_{n=1}^\infty\sideset{}{'}\prod_{\ell,\, m,\, \bar{\ell},\, \bar{m}} \frac{1}{(1-q^m \bar{q}^{\bar{m}}y^\ell \bar{y}^{\bar{\ell}})^{c\left(n\left(m-\ell\right),n\left(\bar{m}-\bar{\ell}\right),\ell-\tfrac{n}{2},\bar{\ell}-\tfrac{n}{2}\right)}}\ .\label{DMVV NSNS sector limit}
\ee
This is actually not the expression with which we should compare the supergravity answer. The reason is that this partition function also counts multi-particle states, whereas we only dealt with single-particle states in supergravity. The transition between the two partition functions is simple, they are related by
\be 
Z_\mathrm{multi}(z|\tau)=\mathrm{exp}\left(\sum_{k=1}^\infty \frac{1}{k} Z_\mathrm{single}(kz|k\tau)\right)\ .
\ee
It is then easy to see that the single-particle version of \eqref{DMVV NSNS sector limit} is
\be 
Z_{\mathrm{NSNS},\, \mathrm{single}}^\infty(z|\tau)=\sum_{n=1}^\infty \sum_{\ell,\, m,\, \bar{\ell},\, \bar{m}} c\left(n(m-\ell),n(\bar{m}-\bar{\ell}),\ell-\tfrac{n}{2},\bar{\ell}-\tfrac{n}{2}\right) q^m \bar{q}^{\bar{m}}y^\ell \bar{y}^{\bar{\ell}}\ .\label{DMVV NSNS sector limit single particle}
\ee
Here, we omitted the prime on the sum, since it simply corresponds to the vacuum in this partition function.
\subsection{The chiral-(anti)chiral spectrum}
We now extract chiral-chiral primary states of \eqref{DMVV NSNS sector limit single particle}. Clearly, only terms with $m=\ell$ and $\bar{m}=\bar{\ell}$ contribute in the sum. Then the sum localizes onto the Ramond ground states of the seed theory. These in turn correspond via spectral flow to chiral-chiral primary states in the seed theory. We use the same trick as in supergravity to determine the chiral-chiral and the chiral-antichiral primary spectrum in one go. For this, we consider the modified supergravity spectrum of K3 and $\mathbb{T}^4$ with the insertions of $\alpha$'s.

\noindent
\medskip
{\bf Enriques surface.} Using the Hodge-diamond \eqref{Enriques surface modified Hodge diamond}, we see that
\vspace*{-.3cm}
\be 
c(0,0,\pm \tfrac{1}{2},\pm \tfrac{1}{2})=1\ , \quad c(0,0,\pm \tfrac{1}{2},\mp \tfrac{1}{2})=\alpha\ , \quad c(0,0,0,0)=10(1+\alpha)\ ,
\ee
and all other ground state coefficients vanish. Thus, the modified K3 supergravity spectrum reads after translating to the supergravity conventions:
\be 
(\mathbf{1},\mathbf{1})_\mathrm{S}^\alpha \oplus (11+10\alpha)(\mathbf{2},\mathbf{2})_\mathrm{S}^\alpha \oplus \alpha(\mathbf{1},\mathbf{3})_\mathrm{S}^\alpha\oplus \bigoplus_{\mathbf{m} \ge 3}\alpha (\mathbf{m},\mathbf{m}\pm 2)_\mathrm{S}^\alpha \oplus  (12+10\alpha) (\mathbf{m},\mathbf{m})_\mathrm{S}^\alpha\ ,
\ee
which is in perfect agreement with \eqref{Enriques surface supergravity spectrum}, up to the aforementioned exceptions at low spin. As a corollary also the chiral-chiral and chiral-antichiral primary spectrum will agree.

\noindent
\medskip
{\bf Hyperelliptic surface.} The Hodge-diamond \eqref{hyperelliptic surface modified Hodge diamond} tells us this time that
\vspace*{-.3cm}
\begin{align} 
c(0,0,\pm \tfrac{1}{2},\pm \tfrac{1}{2})&=1\ ,\quad c(0,0,\pm \tfrac{1}{2},\mp \tfrac{1}{2})=\alpha^{\pm 1}\ , \quad c(0,0,\pm \tfrac{1}{2},0)=1+\alpha^{\mp 1}\ , \nonumber\\
c(0,0,0,\pm \tfrac{1}{2})&=\alpha^{\pm 1}+1\ , \quad c(0,0,0,0)=\alpha^{-1}+2+\alpha\ .
\end{align}
This translates into the following supergravity spectrum from the symmetric orbifold:
\begin{align}
&(\mathbf{1},\mathbf{1})_\mathrm{S}^\alpha \oplus (1+\alpha)(\mathbf{1},\mathbf{2})_\mathrm{S}^\alpha \oplus (\alpha^{-1}+1)(\mathbf{2},\mathbf{1})_\mathrm{S}^\alpha \oplus (\alpha^{-1}+3+\alpha) (\mathbf{2},\mathbf{2})^\alpha_\mathrm{S} \oplus 2(1+\alpha)(\mathbf{2},\mathbf{3})^\alpha_\mathrm{S} \nonumber\\
&\quad \oplus\bigoplus_{m \ge 3} \alpha^\pm (\mathbf{m},\mathbf{m}\pm 2)^\alpha_\mathrm{S} \oplus (1+\alpha^\pm)(\mathbf{m},\mathbf{m}\pm 1)^\alpha_\mathrm{S} \oplus  (\alpha^{-1}+4+\alpha)(\mathbf{m},\mathbf{m})^\alpha_\mathrm{S}\ ,
\end{align}
which is again in agreement with \eqref{hyperelliptic surface supergravity spectrum}, up to low-lying exceptions. Consequently, also the chiral-chiral primary and chiral-antichiral primary spectrum will agree.

Let us mention that we have also applied the techniques developed in \cite{deBoer:1998us, Datta:2017ert} to the present case. We have found that the supergravity elliptic genera of the backgrounds agree with the elliptic genera of the symmetric product orbifold theory. While for the hyperelliptic surfaces, the elliptic genus is vanishing, it equals half of the K3 elliptic genus \eqref{K3 elliptic genus} in the case of the Enriques surface.
\section{Conclusions}
In this paper, we have discussed the conditions imposed by supersymmetry on $\mathrm{AdS}_3$ backgrounds with pure NS-NS flux. We reviewed that supersymmetry implies that the internal manifold is a $\mathrm{U}(1)$-fibration over a conformally balanced manifold. Strengthening the condition to K\"ahler, we were able to give a complete classification of these backgrounds. Moreover, it was relatively easy to identify their dual CFTs.

\bigskip
Several directions for future work seem promising. First, it would be interesting to understand also non-K\"ahler backgrounds on the same level as the K\"ahler backgrounds --- maybe also there a classification could be possible. This would greatly enhance our understanding of $\mathrm{AdS}_3$ backgrounds. Furthermore, one can consider warped products of $\mathrm{AdS}_3$ with $\mathcal{M}_7$, add a non-trivial dilaton profile, or turn on RR-fields. For the latter case, \cite{Kim:2005ez, Donos:2008ug} gives some classification results. Each of these complications adds new interesting ingredients, but the dual CFT will be substantially harder to identify. However, we feel that $\mathcal{N}=(2,2)$ supersymmetry is particularly suited for exploring the landscape of $\mathrm{AdS}_3/\mathrm{CFT}_2$ dualities.

Another exciting direction is black hole counting, particularly for the Enriques surface. The background can be viewed as a near horizon limit of a black hole sitting at a boundary of a five-dimensional space-time. While not a black hole in flat space, one can still perform microscopic state counting. Since the black hole is sitting on the boundary of space-time, the surface of its horizon is precisely half of its original value. This is reflected on the CFT side by the fact that the elliptic genus is half of the K3 value \eqref{K3 elliptic genus}. It would certainly be interesting to explore this in more detail.

Furthermore, one should embed the background into string theory. In particular, it would be interesting to find a suitable D-brane construction, which may provide some insight on how to construct other $\mathcal{N}=(2,2)$ backgrounds.

The symmetric orbifold of the hyperelliptic surface supports a higher spin symmetry. However, it is unknown whether the same holds true for the symmetric orbifold of the Enriques surface at least at special points in the moduli space. For K3, this is possible thanks to free field constructions \cite{Baggio:2015jxa}. In a similar vein, it would be interesting to see whether the corresponding higher spin symmetry can also be realized as (possibly an orbifold of) a coset \cite{Gaberdiel:2014cha}.

Finally, there is still moonshine to be found in the Enriques surface elliptic genus \cite{Eguchi:2013es}. It seems that the Mathieu group $\mathrm{M}_{12}$ acts on the BPS states of this compactification. Hence our construction provides another geometric example of moonshine.
\section*{Acknowledgements}
It is a pleasure to thank Matthias Gaberdiel for guidance in this work and for a careful reading of the manuscript. I also would like to thank Elena Asoni, Shouvik Datta, Andrea Dei, Kevin Ferreira, Anna Karlsson, Christoph Keller, Edward Witten and Ida Zadeh for useful discussions. Furthermore, this manuscript has greatly profited from correspondences with Dario Martelli, Eoin \'O Colg\'ain and Sakura Sch\"afer-Nameki. My research is (partly) supported by the NCCR SwissMAP, funded by the Swiss National Science Foundation. I gratefully acknowledge the hospitality of the Institute for Advanced Study in the final stages of this work.
\appendix
\section{Notations and conventions}\label{app:notations}
We use a mostly plus metric for $\mathrm{AdS}_3 \times \mathcal{M}_7$. Hence $\mathcal{M}_7$ has a standard Riemannian metric. We define a generalized inner product between forms. Assuming $p>q$, it reads:
\be 
(\iota_\alpha \beta)_{a_{q+1} \cdots q_p}\equiv\frac{1}{(p-q)!}\alpha^{a_1\cdots a_q} \beta_{a_1 \cdots a_p}\ .
\ee
For a $p$-form $\alpha$, the Hodge-norm is defined by
\be 
\alpha^2\equiv \iota_\alpha \alpha =\frac{1}{p!} \alpha_{a_1\dots a_p}\alpha^{a_1\dots a_p}\ .
\ee
We have moreover
\be 
\alpha \wedge \star \alpha=\alpha^2\, \mathrm{vol}\ ,
\ee
where $\mathrm{vol}$ is the canonical volume form. For a complex form, we have the natural analog
\be 
\abs{\alpha}^2 \equiv \iota_\alpha \overline{\alpha}\ .
\ee
The Hodge dual in $n$ dimensions and Euclidean signature of a $k$-form is defined to be
\be 
(\star \,\alpha)_{a_1 \cdots a_{n-k}}=\frac{1}{k!}\epsilon_{b_1 \cdots b_k b_{k+1} \cdots b_n} \alpha^{b_1 \cdots b_k}\ .
\ee
\section{Some properties of complex surfaces} \label{app:surfaces}
In this appendix, we collect some interesting and useful properties of the complex surfaces we use in the main text, namely the four-torus $\mathbb{T}^4$, $\mathrm{K3}$, the Enriques surface $\mathrm{ES}$ and the hyperelliptic surfaces $\mathrm{HS}$.

All of these surfaces are projective and therefore K\"ahler. They are furthermore distinguished among other complex surfaces, since their first Chern class vanishes in real cohomology, i.e.~it is a torsion element in integer cohomology.
This suffices for Yau's Theorem \cite{Yau:1977ms} to hold and consequently these surfaces support a Ricci-flat metric. It is furthermore possible to use them as target spaces for $\mathcal{N}=(2,2)$ sigma-models, since there is no axial anomaly. 

We should note that there are other complex surfaces with vanishing first Chern class in real cohomology. These are the primary and secondary Kodaira surfaces. However, these are not algebraic and hence not K\"ahler, so they are unsuitable for our purposes.
\subsection{$\mathbb{T}^4$}
$\mathbb{T}^4$ is certainly the most explicit of the four surfaces. In particular the Ricci-flat metric is the canonical metric inherited from $\mathbb{C}^2$, when thinking of $\mathbb{T}^4$ as a quotient thereof. The Hodge-diamond reads
\be 
\begin{tabular}{ccccc}
& & 1 & & \\
& 2 & & 2 & \\
1 & & 4 & & 1 \\
& 2 & & 2 & \\
& & 1 & &
\end{tabular}
\ee
and the cohomology ring is the exterior algebra over four generators -- two of degree $(1,0)$ and two of the degree $(0,1)$. To determine the action of group actions on the cohomology, it hence suffices to determine the action on these four generators. The Euler-characteristic, the signature and all other genera of the surface vanish. The canonical bundle is trivial. $\mathbb{T}^4$ is a spin manifold.
\subsection{$\mathrm{K3}$}
$\mathrm{K3}$ is the unique simply-connected Calabi-Yau surface. It can be realized as various orbifolds of $\mathbb{T}^4$. However away from these orbifold points, the Ricci-flat metric is not explicitly known, but exists thanks to Yau's Theorem. The Hodge-diamond reads
\be 
\begin{tabular}{ccccc}
& & 1 & & \\
& 0 & & 0 & \\
1 & & 20 & & 1 \\
& 0 & & 0 & \\
& & 1 & &
\end{tabular}\ .
\ee
The Euler-characteristic is 24, while the signature is $-16$ --- the intersection lattice is the unimodular lattice $\mathrm{II}_{3,19}$. 
The holonomy group equals $\mathrm{SU}(2)$. The canonical bundle is again trivial. K3 is also a spin manifold.
The elliptic genus of string theory is non-vanishing and equals
\be 
\mathcal{Z}_\mathrm{K3}(z|\tau)=8 \left(\frac{\theta_2(z|\tau)^2}{\theta_2(\tau)^2}+\frac{\theta_3(z|\tau)^2}{\theta_3(\tau)^2}+\frac{\theta_4(z|\tau)^2}{\theta_4(\tau)^2}\right)\ . \label{K3 elliptic genus}
\ee
\subsection{$\mathrm{HS}$}
Hyperelliptic surfaces are finite quotients of tori --- we gave an overview of the different possibilities in Table~\ref{tab:hyperelliptic surfaces}. The Hodge-diamond reads for all possibilities
\be 
\begin{tabular}{ccccc}
& & 1 & & \\
& 1 & & 1 & \\
0& & 2 & &0 \\
& 1 & & 1 & \\
& & 1 & &
\end{tabular}\ .
\ee
Hyperelliptic surfaces are elliptic fibrations over elliptic curves. For this reason, they are also called bi-elliptic surfaces.
The Euler-characteristic vanishes in all cases, as does the signature. The holonomy group is $\mathbb{Z}_n$, of which the generator is a rotation by an angle of $\tfrac{2\pi}{n}$. Here, $n=2$, $3$, $4$ and $6$ for type a, b, c and d, respectively. The canonical bundle is a torsion bundle, i.e.~it is not trivial but its $n$-th power is. Finally, hyperelliptic surfaces are not spin manifolds.
\subsection{$\mathrm{ES}$}
Enriques surfaces can be realized as $\mathbb{Z}_2$-quotients of $\mathrm{K3}$ surfaces. They have Hodge-diamond
\be 
\begin{tabular}{ccccc}
& & 1 & & \\
& 0 & &0 & \\
0& & 10 & &0 \\
& 0 & & 0 & \\
& & 1 & &
\end{tabular}\ .
\ee
The Euler-characteristic is 12, the signature is $-8$. The intersection lattice is the unimodular lattice $\mathrm{II}_{1,9}$. The canonical bundle is a torsion bundle of order two. The holonomy group is a semidirect product $\mathrm{SU}(2) \rtimes \mathbb{Z}_2$. Finally, Enriques surfaces are not spin manifolds. The string theory elliptic genus is half of the K3 elliptic genus. 
\section{Modifying $\mathcal{N}=4$ multiplets} \label{app:N4N2}
To determine the BPS and supergravity spectrum in the main text, we used the underlying $\mathcal{N}=4$ multiplet structure of the compactification. In this appendix, we provide some details of the modifications of the $\mathcal{N}=4$ multiplet structure we used.

We first pick an $\mathcal{N}=2$ subalgebra inside the $\mathcal{N}=4$ algebra of which the corresponding supercharges will be denoted by $G^+_r$ and $G^-_r$. The remaining two supercharges will be denoted by $\widetilde{G}^+_r$ and $\widetilde{G}^-_r$. They are not invariant under the quotient we are performing --- they have eigenvalues $\alpha$ and $\alpha^{-1}$, respectively. Here, $\alpha$ denotes a unit root of the same order as the group by which we are performing the quotient. Similarly, the Cartan-element of the $\mathfrak{su}(2)$-current algebra is invariant under the quotient, while the two raising and lowering operators $J^\pm_m$ pick up the eigenvalues $\alpha^\pm$.

Let us denote by $\chi_\ell(y)$ an $\mathfrak{su}(2)$ character of spin $\ell$. We further denote by $\chi_\ell^\alpha(y)$ an $\mathfrak{su}(2)$-character twisted by $\alpha$:
\be 
\chi_\ell^\alpha(y)=\sum_{\genfrac{}{}{0pt}{}{j=-\ell}{j+\ell \in \mathbb{Z}}}^\ell \alpha^{\ell-j} y^j\ .
\ee
The corresponding multiplet will be denoted by $(\mathbf{m})^\alpha$ in the main text, where $\mathbf{m}=2\ell+1$. When combining left- with right-movers, we write $(\mathbf{m},\mathbf{n})^\alpha$ for the twisted multiplet. One has to pay attention that one has to use $\alpha$ for the left-movers and $\alpha^{-1}$ for the right-movers, i.e.~we have
\be 
(\mathbf{m},\mathbf{n})^\alpha \cong (\mathbf{m})^\alpha \otimes (\mathbf{n})^{\alpha^{-1}}\ .
\ee
It is simple to write down a short $\mathcal{N}=4$ character of the global $\mathfrak{su}(1,1|2)$-algebra twisted by $\alpha$:
\be 
\chi_\ell^{\mathcal{N}=4,\, \alpha}(q,y)=\frac{q^\ell}{1-q}\left(\chi^\alpha_\ell(y)-q^{\tfrac{1}{2}}(1+\alpha)\chi^\alpha_{\ell-\frac{1}{2}}(y)+q\alpha \chi^\alpha_{\ell-1}(y)\right)\ .
\ee
Here, we inserted a $(-1)^\mathrm{F}$ in the definition of the character. An $\mathcal{N}=2$ character is by definition invariant under the projection and reads
\begin{align} 
\chi^{\mathcal{N}=2}_{j,h}(q,y)=\frac{q^h}{1-q}\begin{cases}
y^j+q^{\frac{1}{2}}y^{j-\frac{1}{2}} & \qquad j=h\ , \\
y^j+q^{\frac{1}{2}}y^{j+\frac{1}{2}} & \qquad j=-h\ , \\
y^j+q^{\frac{1}{2}}(y^{j-\frac{1}{2}}+y^{j+\frac{1}{2}})+q y^j & \qquad -h<j<h\ ,
\end{cases}
\end{align}
where the three cases correspond to chiral primary, anti-chiral primary and long representations, respectively. It is then simple to check that we have the following decomposition:
\be 
\chi_\ell^{\mathcal{N}=4,\, \alpha}(q,y)=\sum_{\genfrac{}{}{0pt}{}{j=-\ell}{j+\ell \in \mathbb{Z}}}^\ell \alpha^{\ell-j} \chi^{\mathcal{N}=2}_{j,\ell}(q,y)\ ,
\ee
where the bottom and top components of the sum are short.

\end{document}